# Beam Pointing of Relativistic High-order Harmonics Genrated on a Nonuniform Pre-plasma

Chaoneng Wu[1,2], Yiming Xu[2,3], Andre Kalouguine[4], Jaismeen Kaur[4], Antoine Cavagna[4], Zuoye Liu[3], Rodrigo Lopez-Martens[4], Cangtao Zhou[1,2], Philippe Zeitoun[4], Stefan Haessler[4,*], Lu Li[2,*]

[1] *College of Physics and Optoelectronic Engineering, Shenzhen University, Shenzhen 518060, China*
[2] *Shenzhen Key Laboratory of Ultraintense Laser and Advanced Material Technology, Center for Intense Laser Application Technology, and College of Engineering Physics, Shenzhen Technology University, Shenzhen 518118, China*
[3] *Frontiers Science Center for Rare Isotopes and School of Nuclear Science and Technology, Lanzhou University, 730000 Lanzhou, China*
[4] *Laboratoire d'Optique Appliquée, Institut Polytechnique de Paris, ENSTA Paris, Ecole Polytechnique, CNRS, Palaiseau, France*
*stefan.haessler@ensta-paris.fr*

*lilu@sztu.edu.cn*

**The use of tunable pre-pulse is a common technique to enhance the high-order harmonic generation from surface plasma. The shape and dynamic of the electron density, the degree of ionization and its rate, and the plasma heating are influenced by the pre-pulse properties. Non-uniform pre-pulse could cause a spatially varying density map to the pre-plasma region, which serves as the spectrally up-conversion and reflection surface. The corresponding geometrical feature and plasma nature under laser field will affect the harmonic emission properties. In this study, the variation in harmonic beam pointing due to the electron density shape was investigated. Particle-in-cell simulations demonstrated that both plasma hydrodynamics and geometrical optical effect induce the deviation of harmonic beam from specular reflection. This research contributes to the understanding of the surface plasma dynamics during high harmonic generation process.**

The advancement of laser technology has facilitated the generation of relativistic electromagnetic fields capable of generating extreme ultraviolet (EUV) radiation via laser-driven plasma surfaces. The surface high-order harmonic generation (SHHG) has the advantage on avoiding the driving field intensity threshold for gas medium [1] and high conversion efficiency [2, 3, 4], allowing to deliver ultra-intense attosecond pulses at EUV region. Considerable research has been dedicated to elucidating the mechanisms of SHHG [3, 4], including coherent wake emission (CWE), relativistic oscillating mirror (ROM), and coherent synchrotron emission (CSE). In the past fifteen years, the SHHG source development has evolved from physical understanding to characterization and field manipulation. Spectral, spatial, temporal characterizations and modulation via laser-plasma interaction control are crucial for the source application. Notably, ptychography [5] and the Spectral Phase Interferometry for Direct Electric-Field Reconstruction (SPIDER) [6] methods have been proposed to accurately characterize both the spatial and spectral phases of the SHHG source. Concerning harmonic source modulation, conversion efficiency optimization is closely related with the temporal modulation techniques. These include two-color field enhancement of harmonic spectra [7, 8] and optimization using tailored multi-color driving lasers [9]. The underlying physics is the reaction of plasma surface to the laser field, which plays an important role in controlling harmonic qualities [10]. The SHHG process is particularly sensitive to electron density gradient at the plasma-vacuum boundary [11]. Recent works show that the spatial density distribution can also affect the SHHG properties. The relativistic SHHG emission can be influenced by the plasma denting effect, which significantly alters the critical surface [12, 13]. Methods proposed to compensate for spatial degradations include plasma truncation with an extra pre-pulse [14], slightly moving the target out of focus for near-field wave front compensation [15] and using a convex target to achieve a flat surface despite plasma denting-induced concavity [16]. CWE control was also studied via pre-plasma distribution manipulation [17, 18].

Experimental approaches to create a controllable steep pre-plasma profile have been widely developed. Typically, a small fraction of the laser energy is diverted from the main beam using a holey or pick-up mirror along the transmission line prior to reaching the focal spot [19]. The small pre-pulse beam and the larger main beam then follow the same optical path, ensuring the main beam to be accurately positioned at the plasma-vacuum boundary. Essentially, the harmonic signal is collected in the specular reflection direction, hundreds of millimeters distant from the laser-plasma interaction point, within the far-field region of the harmonic beam. The harmonic signal typically exhibits a divergence angle ranging from tens to 100mrad [5,11,12,19]. For detection metrology and applications, controlling the beam pointing is crucial.

The harmonic characteristics, including directionality, have been the subject of theoretical and experimental studies over the past twenty years. However, a common limitation of current theories is the assumption of an 1D plasma electron density distribution, which is typically modeled as a transversely uniform exponential decay profile. Practically, the pre-pulse induced plasma electron density can exhibit significant lateral spatial, gradients leading to an initial reflecting surface with a tilt. When the main pulse interacts with this tilted surface, the harmonic beam deviates from the specular direction. To demonstrate this effect, a Gaussian pre-plasma density profile along target surface is considered. The pre-pulse will induce a displacement of the pre-plasma, and in turn the main pulse will not be centered on it. Even under ideal optical conditions, the thermal expansion of the pre-plasma contributes to the angular deviation, as depicted in Fig. 1(a). The target is placed on x-z plane, while y axis is perpendicular to the target. Considering the laser rays lie on x-y plane for simplicity, the problem can be solved on the x-y plane. Note that we can simply extend the following theory to obtain comprehensive 3D results. In figure 1, the pre-pulse and main pulses are illustrated by the lighter and darker red cones, respectively. The plasma region, where the density exceeds the critical density ($n_e = n_c \cos^2\theta$), is highlighted in blue The shape of pre-plasma is caused by the uneven expansion speed of the exponential density ramp, which is known to increase with the pre-pulse fluence [19,20,21]. According to the theory of pre-plasma scale along target surface, a Gaussian approximation is given by

$$\boldsymbol{n_e(x,y) = n_0\ e^{-\frac{y}{L(x)}-1}, L(x) = L_0\ e^{-\frac{x^2}{w_{pre}^2}}}, \quad (1)$$

where $n_e(x,y)$ is the electron density profile [21], $n_0$ is the peak density reached in the solid target bulk, x is the distance along solid surface to the beam center, and y is the distance along the target normal from the initial solid surface before thermal expansion, $w_{pre}$ is the 1/e-radius of the pre-plasma gradient scale-length profile.

To elucidate the impact of the Gaussian pre-pulse on the reflected beam, we initially assume the critical surface as a static and perfect reflector. Upon the main pulse striking the pre-plasma at an oblique angle ($L \neq 0$), the interaction point between the laser and plasma occurs where the Gaussian-shaped critical surface intersects with the central axis of the main pulse's optical path. Using the scenario without a pre-plasma ($L_0 = 0$) as a benchmark to measure beam deviation caused by geometric tilt, the EUV beam propagates along the specular reflection, indicated by the purple dashed lines. To address the beam deviation when a pre-plasma is present ($L_0 \neq 0$), the location of the reflection, ($x_r, y_r$), is determined by calculating the intersection between the critical surface $y_1(x)$ and the light ray path equation (Eq.) $y_2(x)$, which are represented as:

$$\boldsymbol{y_1(x) = L_0 \exp\left[-\frac{x^2}{w_{pre}^2}\right] \ln\left(\frac{n_0/e}{n_c \cos^2\theta}\right), y_2(x) = \frac{x-x_0}{\tan\theta}}, \quad (2)$$

Here, $e$ is Euler's number (according to Eq. 1, the density at the original solid-vacuum boundary, y=0, remains constant at $n_e = n_0/e$ over the whole laser heating area),

The inclination of the reflecting surface due to the pre-plasma expansion, as depicted in Fig. 1(a), causes a deviation of the reflected beam from the expected specular direction by the angle

$$\boldsymbol{\alpha = 2\arctan\left.\frac{dy_1(x)}{dx}\right|_{x=x_r} \approx 2\ln\left(\frac{n_0/e}{n_c \cos^2\theta}\right)\delta(x_r)}, \quad (3)$$

where the slope $\delta(x_r)$ is defined as $\delta(x) = dL/dx|_{x=x_r}$ It is a more universal indicator for the pre-plasma tilt than the specific $L$ and $w_{pre}$ in Eq.1, and for a Gaussian pre-plasma can be written as:

$$\boldsymbol{\delta = -2x_r L(x_r)/w_{pre}^2}, \quad (4)$$

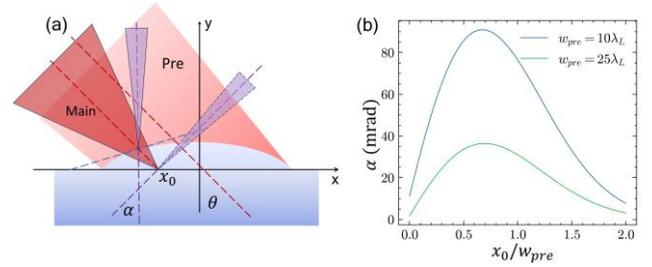

Fig. 1. (a) Schematic for the origin of pre-plasma tilt: the interaction between a Gaussian distributed pre-plasma and the main pulse. The light red beam represents the pre-pulse, and the dark red beam indicates the main pulse. For clarity in the figure, the main pulse has been exaggerated in size. The purple lines signify the reflection directions with and without pre-plasma. (b) angle deviation induced by the geometrical tilt at different focal positions.

According to the focal position of main pulse $x_0$ and the density distribution, $L$ and $\delta$ can be regarded as two independent indicators for describing the pre-plasma condition in the main pulse focus. Fig. 1(b) shows the deviation angle $\alpha$ at different incident positions for two different pre-plasma conditions: a narrow pre-plasma profile with $w_{pre} = 10\lambda_L$ and a normal one with $w_{pre} = 25\lambda_L$, $\theta = \pi/4$ and $L = 0.1\lambda_L$ are the same in these two cases. Assuming a bulk plasma density such that $n_0/e = 100n_c$, which is close to the case of a fully ionized $SiO_2$, the deviation angle $\alpha$ reaches tens of milliradians at $x_r = \frac{w_{pre}}{\sqrt{2}}$. This is very similar to typically observed SHHG beam divergences [5,11,12,19] and can thus be obviously observed in experiments. While the deviation angle at $x_0 = 0$ remains fairly small ($\alpha \approx 10$ mrad) even with the narrow pre-plasma profile, for longer scale lengths $L > 0.1\lambda_L$, significant beam deviations may occur. The transverse scale-length variation $\delta$ at the flank of the pre-plasma, with a maximum value of $\delta \approx 8 \times 10^{-3}$ for the narrow pre-plasma profile considered above, is identified as the origin of an effective surface tilt, deviating the reflected beam.

The above model describes the geometrical angle deviation. However, upon the incidence of a relativistic, tightly focused short pulse, the reflection does not happen at a fixed surface but is caused by a dynamic compressed electron sheet driven around the critical density region. This compression modifies the reflection spot relative to the original surface position, introducing a correction to the geometrical deviation. To assess the impact of the plasma effect on the directionality of the harmonic source, particle-in-cell (PIC) simulations were conducted with the WarpX code [22]. They show that the actual tilt angle observed on the reflected pulse is reduced compared to the geometrical tilt angle, suggesting a correction coefficient <1 should be applied to the right-hand side of Eq. (3). Once this correction is known, the easily measurable modulation of the SHHG beam direction directly indicates the critical surface tilt, which in turn can be employed to characterize the plasma expansion process within a pump-probe like scheme, where the pump is the pre-pulse and the probe is the SHHG-driving pulse. This approach can provide valuable insights into the behavior of the plasma under the influence of the driving laser field.

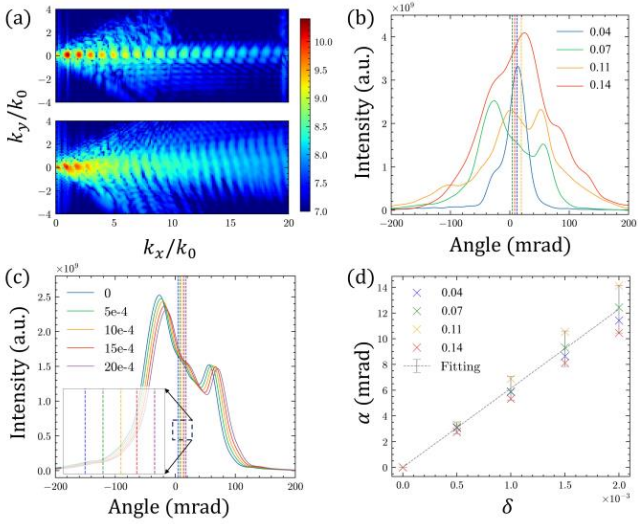

Fig. 2 Simulation results for different scale length L and decay slope $\delta$. (a) 2D spectrum of $L = 0.04$ and $0.14$, (b) lineouts of far-field angular spectrum map driven by different scale lengths, the centroid of harmonic spectrum is shown in dashed lines, (c) the spectrum drift caused by increasing $\delta$, the centroid of spectrum is shown in dashed lines and zoomed in in the inset figure. (d) the simulated angular drift by different decay slope, the fitting linear relationship is shown with error bar, which indicates for $d\alpha/d\delta = 6.1 \pm 0.9$.

The results of the PIC simulations are illustrated in Fig. 2(a). A p-polarized laser is injected into the simulation box, with a beam waist of $w_0 = 1.5\lambda_L$ and a duration of $\tau = 2T_0$. Here, $\lambda_L = 800$ nm and $T_0 = \lambda_L/c$ represent the fundamental laser wavelength and period, respectively. The normalized vector potential is $a_0 = 2.5$, and the incidence angle is chosen as $\theta = 45°$. The target consists of a fully ionized carbon slab with a thickness of $2\lambda_L$, with a maximum density of $n_{max} = 100n_c$. This maximum density remains constant at the initial slab surface and exponentially decays with rate $L$ that varies along the slab surface. Since the plasma denting depth in the conditions considered here remains much smaller than the distance from the critical density surface to the $100n_c$-surface, this is equivalent to the situation considered for the geometrical model above, where Eq.1 fixed the plasma density at $100n_c$ at the initial vacuum-plasma interface and let it exponentially decay towards vacuum with rate $L$. The simulation grid size is set to $\lambda_L/512$, and 100 particles per cell are used for resolving up to the 30$^{th}$ order harmonic signals. With a slope $\delta$, the scale length $L(x)$ of the pre-plasma varies with $x$ according to $L = L_c + \delta x$, where $L_c$ is the central scale length at $x = 0$. Typical results with $L_c/\lambda_L = 0.04, 0.07, 0.11, 0.14$ and $10^4\delta = 5, 10, 15, 20$ are shown in Fig. 2. The beam drift was quantified through the angular distribution of the simulated harmonics. A near-field-to-far-field transformation was applied to project the reflected harmonic beam from the simulation domain into the far-field. Subsequently, the drift angle $\alpha$ was determined by the displacement from the specular reflection direction of the centroid of the high-order harmonic ($n > 10$). The simulated near-field spectrum captured at simulation box boundary for $L_c = 0.07$, $\delta = 0$ and $L_c = 0.14$, $\delta = 0$ are shown in Fig. 2(a) as an example, where $k_1$ is along the specular reflection direction. The corresponding far-field angular distributions for various $L_c$ values with $\delta = 0$, were calculated by integrating the spectral intensity along each far-field angular direction and are shown in Figure 2(b). The centroids in the far-field angular distribution are shown by dashed lines, which indicate the angular drift induced by the scale-length variation. The inherent angle $\alpha_0$ for the uniform target with $\delta = 0$ was found to be 12 mrad, 4 mrad, 19 mrad and 8 mrad for the scale lengths $L_c/\lambda_L = 0.04, 0.07, 0.11, 0.14$, respectively. In this scenario, few-mrad level angular drifts from the specular direction are observed, which are small compared to the overall angular divergence and are dominantly caused by the morphing beam shape, without a clear dependence on $\delta$. Note that a very tightly focused driving bean is adopted in this work, and when it comes to the typical experimental results with an angular divergence of ~50 mrad, the beam drift due to variations in $L_c$ can be expected to be less than 5 mrad.

While the scale length $L_c$ has a minimal impact on the beam directionality, the decay rate $\delta$ influences the angular drift. With $L_c$ set to $0.07\lambda_L$, the angular intensity distributions for various $\delta$ values are depicted in Figure 2(c). The angular profile shifts monotonously without changing its general shape, demonstrating a linear relationship between the angular drift $\alpha$ (with respect to $\delta = 0$) and the slope $\delta$, consistent with the prediction from Eq. (3). However, Figure 2d reveals, across different scale lengths $L_c$, a linear slope $k_{PIC} = d\alpha/d\delta = 6.1 \pm 0.9$, which is below the value the value derived from Eq. (3): $k_1 = 10.6$. This discrepancy confirms our expectation of a correction factor <1 arising from the plasma effect that affects the geometrical angular deviation.

To elucidate the plasma correction, an analysis of the electron dynamics in the pre-plasma was conducted. The SHHG process can be divided into three distinct phases based on the motion of the reflecting surface: compression phase, acceleration phase, and dephasing phase. According to the plasma denting theory [12], electrons in the pre-plasma reach an equilibrium position at the end of the compression phase. It can be considered as an indicator for the emission depth. In the case of short driving pulses, the effect of the positive ion background motion can be neglected. The equilibrium depth that the reflecting surface ultimately reaches, relative to the geometrical surface position [12], is given as:

$$y_e = -L \ln\left[1 + \frac{\lambda_L a_0(1 + \sin\theta)}{\pi L}\frac{1}{\cos^2\theta}\right], \quad (5)$$

Noted that $y_1(x)$ in Eq. 2 is not limited to a Gaussian distribution, every function has its own geometrical decay slope, and thus the geometrical linear factor can be calculated from Equation (3). The plasma denting effect can thus provide a correction to the reflection position via $dy_e(x)/dx$. Since $y_e$ is approximately linearly related to $L$ while the main beam is small compared to the pre-plasma, it results in a nearly constant correction to the angle deviation $\alpha \approx 2\, dy_1/dx$, which holds true while $\alpha$ is small in our scenario.

In physical terms, plasma denting flattens the slope $\delta(x_r)$ because a larger scale length facilitates the pushing of the reflecting electron sheet further into the dense plasma region. When the laser swipes over the plasma surface in Fig. 3(a), the equilibrium depth difference between the tilt case with $\delta \neq 0$ and the uniform case with $\delta = 0$ can be regarded as the reason for the beam drift. The three panels show snapshots at different instants spaced by one laser cycle, where the center panel gives the density map when the electron sheet reaches $x = 0$. Vertical lineouts of the electron density at the three positions marked in Fig. 3(a) are plotted in Fig. 3(b). To exhibit how $\delta$ tilts the reflective surface, the lineouts at the same instants and positions are shown for $\delta = 0$ as a reference. We observe no difference at the beam center (position 2), and the sheet in the flat case ($\delta = 0$) starts from a shallower position and ends at a deeper position, opposite to the tilt case ($\delta = 0.002$). The reflective-surface dynamics around the equilibrium position are depicted in Fig. 3(c). Using the equilibrium depth at $x = 0$ (position 2 in Fig. 3(a)) as a reference, if the equilibrium depth at position 1 was calculated by only considering the geometrical tilt of the critical surface, it should be located at the blue dotted line. If the correction by the plasma effect from $dy_e(x)/dx$ is considered, the equilibrium depth is predicted perfectly. That means that our combined theory with the geometrical tilt and the plasma effect can be used to calculate the equilibrium depth for different spatial positions. The relative difference between the equilibrium position of a flat and a tilted surface can be used to evaluate the correctness of Eq. 5, shown in Fig. 3(d). The combined theory predicts a reduced tilt of the effective reflecting plasma surface compared from the purely geometrical model. The slope was evaluated by considering the depth difference between positions 2 and 3. Good agreement is shown between the combined theory and PIC-simulation results. Quantitatively, the angle deviation induced by the plasma denting correction can be expressed here for our simulation parameters as:

$$k_e = \frac{d}{d\delta}\left(2 \arctan \frac{dy_e}{dx}\right) \approx 4.8, \quad (6)$$

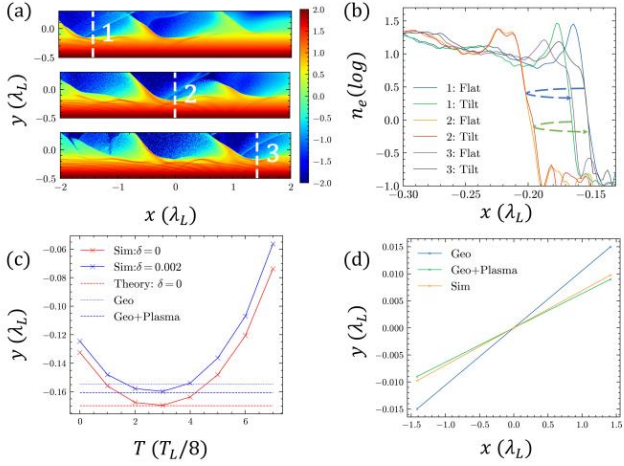

Fig. 3. Correction of decay slope on the geometrical profile by the plasma effect. (a) density maps driven by laser at different instants for $\delta = 0.002$ and (b) the corresponding lineouts at the equilibrium positions for $\delta = 0$ and 0.002, the dashed lines shows the equilibrium surface dynamics along $x$ when laser swipes along target, blue represents for the flat target and green represents for the tilt one, (c) the reflecting surface motion around the end of compression phase for position 1, (d) denting depth difference between $\delta = 0$ and 0.002.

Note that $k_e$ varies with $a_0$, and here the result at $x = \sqrt{2}\lambda_L$ was used as an averaged value. The combined angular drift is thus $\alpha = (k_1 - k_e)\delta = 5.8\delta$, which fits well the simulation results of $k_{PIC} = 6.1 \pm 0.9$. Note that when a long driving pulse is used, a further correction should be considered to account for ion motion. The detailed correction from scale-length and ion motion is out of scope of this article, and it will be studied in our next work.

In conclusion, this study has thoroughly examined the effects of SHHG beam drifting for various pre-plasma distributions. For SHHG driven by few-cycle laser, the deviation of the harmonic emission angle from the specular direction is primarily governed by the slope of the critical surface. The theory and observed phenomena indicate that the drifting of the harmonic beam can serve as a probe for the critical surface condition at the main pulse focal spot. When a high-contrast laser system is utilized, the geometrical tilt dynamics of the pre-plasma can be continuously measured with varying pre-pulse delays. Furthermore, SHHG beam drifting as well as the far-field wave front information carries information on the relativistic plasma at micrometer and femtosecond to picosecond scales.

**Funding.** National Natural Science Foundation of China (Grant No. 12205203), LASERLAB-Europe 5 (LOA0024038);

**Acknowledgments.** This research used the open-source particle-in-cell code WarpX https://github.com/ECP-WarpX/WarpX. Primary WarpX contributors are with LBNL, LLNL, CEA-LIDYL, SLAC, DESY, CERN, and TAE Technologies. We acknowledge all WarpX contributors.

**Disclosures**. The authors declare no conflicts of interest.

**Data availability.** Data underlying the results presented in this paper are not publicly available at this time but may be obtained from the authors upon reasonable request.